\documentclass[12pt]{article}
\usepackage{graphicx,wrapfig}
\usepackage{floatflt}

\newcommand\pubnumber{NuPhys2016-Riccio}
\newcommand\pubdate{\today}
\def\napoli{INFN Sezione di Napoli and Universit\'a di Napoli Federico II, Dipartimento di Fisica "E. Pancini",
	 Napoli, Italy}

\def\Title#1{\begin{center} {\Large #1 } \end{center}}
\def\Author#1{\begin{center}{ \sc #1} \end{center}}
\def\Address#1{\begin{center}{ \it #1} \end{center}}

\newcommand\pubblock{\rightline{\begin{tabular}{l} \pubnumber\\
         \pubdate  \end{tabular}}}
\newenvironment{Abstract}{\begin{quotation}  }{\end{quotation}}
\newenvironment{Presented}{\begin{quotation} \begin{center} 
             PRESENTED AT\end{center}\bigskip 
      \begin{center}\begin{large}}{\end{large}\end{center} \end{quotation}}

\begin{document}
\begin{titlepage}
\pubblock

\vfill
\Title{Muon antineutrino charged-current cross sections without pions in the final state at T2K}
\vfill
\Author{Ciro Riccio\footnote{riccioc@na.infn.it}, \\for the T2K Collaboration}
\Address{\napoli}
\vfill

\begin{Abstract}
T2K (Tokai to Kamioka) is a long-baseline neutrino oscillation experiment located in Japan and designed to measure neutrino flavor oscillation using an off-axis neutrino beam. Data collected recently with an antineutrino beam allows T2K to measure cross sections for antineutrinos at an energy around 600 MeV using the off-axis near detector. These measurements, along with the analogous for neutrinos, are vital inputs to neutrino oscillation analyses and their interpretation. In this work preliminary results on the simultaneous extraction of the muon neutrino and antineutrino charged-current cross sections without pions in the final state is presented. The two cross sections will be measured as a function of muon kinematic allowing to evaluate the sum, difference and asymmetry between the two cross sections. These results are useful for comparison and tuning of theoretical models of nuclear effects such as multinucleon interactions (also know as 2 particle-2 hole processes).
\end{Abstract}
\vfill
\begin{Presented}
NuPhys2016, Prospects in Neutrino Physics
Barbican Centre, London, UK,  December 12--14, 2016
\end{Presented}

\vfill
\end{titlepage}
\def\thefootnote{\fnsymbol{footnote}}
\setcounter{footnote}{0}

\section{Introduction}

Modern neutrino oscillation experiments use neutrino beams with energies around 1 GeV, where the main interaction channel is the charged current quasi-elastic scattering (CCQE). This bring to the necessity of a correct modelling of this process. The K2K experiment~\cite{k2k}, MiniBoone experiment~\cite{mboone} toghether with other experiments, using carbon as target, measured a kinematic distributions of the outgoing muons in CCQE interactions not consistent with the prediction of the Relativistic Fermi Gas (RFG) nuclear model~\cite{rfg-smith72,rfg-smith75}. Many theoretical models have been proposed to describe the nuclear effects in (anti-)neutrino scattering~\cite{martini1,amaro,nieves2011}. They predict the existence of multinucleon emission (np-nh) that produces an increase in the CCQE-like cross section, since the additional outgoing nucleons in the final state could remain undetected, mimicking a QE interaction. In reference~\cite{martini2} it has been pointed out that the multinucleon emission can contribute in a different way in $\nu_\mu$ and $\bar{\nu}_\mu$ interactions. This gives the possibility to isolate the np-nh contribution looking at different linear combinations of the neutrino and antineutrino cross-sections. During the last two years T2K collected data in antineutrino mode opening the possibility to study the antineutrino cross section.

\section{The near detector of the T2K experiment}

The T2K off-axis near detecor (ND280) is a fully magnetized particle tracking detector located 280 m downstream of the beam target and 2.5 degrees off the neutrino beam center. Placed inside the refurbished UA1/NOMAD magnet which provide a magnetic field of 0.2 T, it consists of a $\pi^0$ detector (P$\O$D), two Fine Grained Detectors (FGDs)
\begin{wrapfigure}[11]{l}{5.2cm}
	\vspace{-2ex}%
	\includegraphics[width=0.3\textwidth]{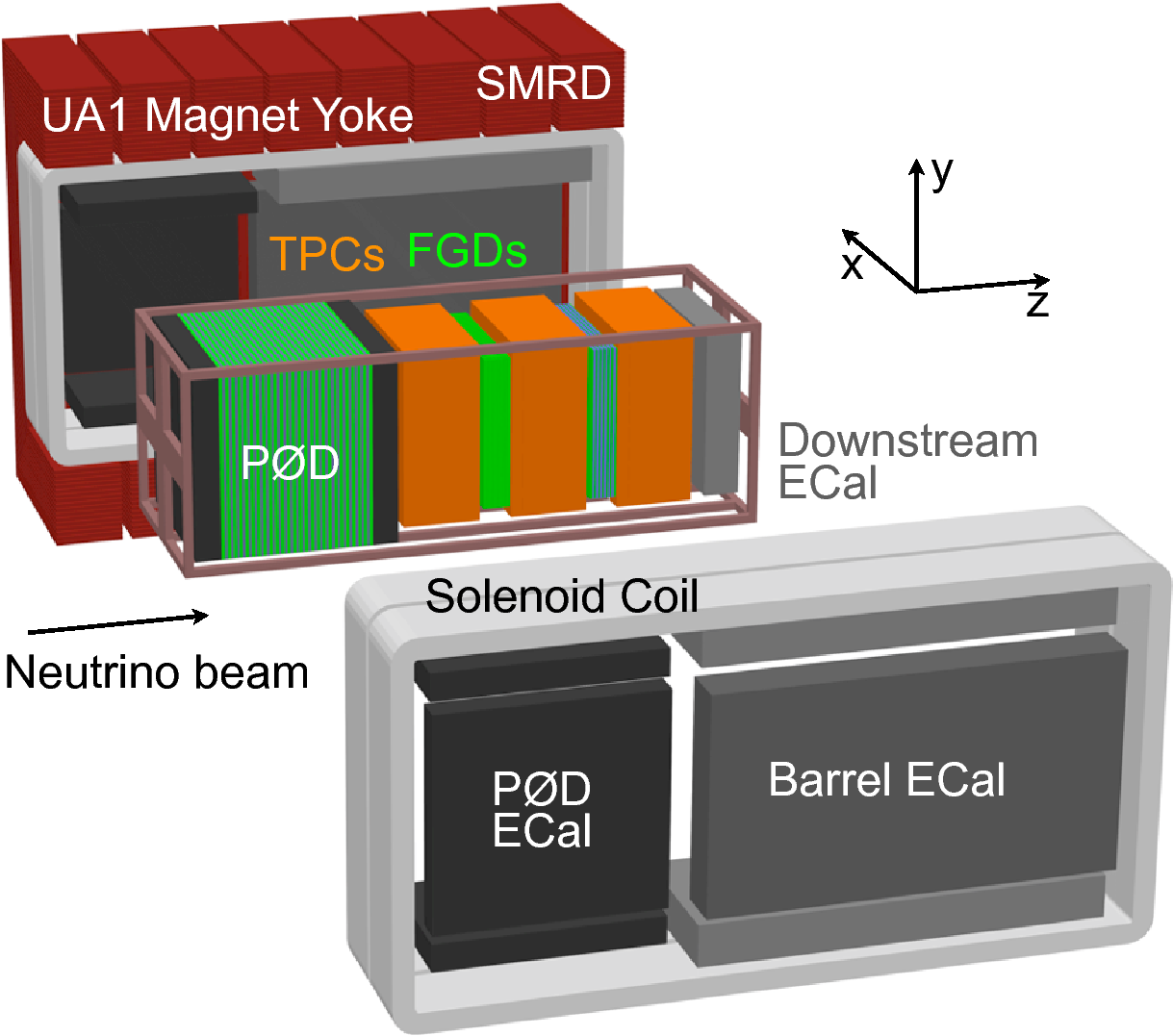}
	\caption{View of ND280~\cite{t2knim}.}\label{fig:nd280biggerlabel}
\end{wrapfigure}
 and three gas Time Projection Chambers (TPCs), surrounded by an electromagnetic calorimeter (ECal) as shown in Fig.~\ref{fig:nd280biggerlabel}. The FGDs, composed of finely segmented scintillator bars, serve as targets for neutrino interactions and provide tracking capabilities which allow measurements of short tracks. Upstream FGD (FGD1) contains only scintillator bars arranged in x and y directions with respect to the beam direction, while in downstream FGD (FGD2) there are additional water layers placed in between scintillator layers. The two FGDs are installed amongst the TPCs designed for 3D track reconstruction, particle identification and determination of momentum and charge. The ECals are sampling calorimeters consisting of layers of plastic scintillator and layers of lead. They allow reconstruction of tracks and showers in order to further distinguish between muons, electrons and pions.

\section{Analysis strategy and preliminary results }

The cross sections are simultaneously extracted using a binned likelihood fit in muon kinematic variables (similar to the method used in Ref.~\cite{t2kcc0pi}). In this work only preliminary results usign only the Monte Carlo (MC) prediction are shown. The MC generator used is NEUT 5.3.2 which uses the Llewellyn-Smith formalism~\cite{llewellyn-smith} to model CCQE interactions, the Spectral Function (SF) model by Behnar et al.~\cite{benhar} as nuclear model and a nominal axial mass $M_A^{QE}$ set to 1.21 GeV.  Multinucleon interactions are modeled with an implementation of the model by Nieves et al.~\cite{nieves2011}.
\begin{figure}[htbp]
	\centering
	\includegraphics[width=.32\textwidth]{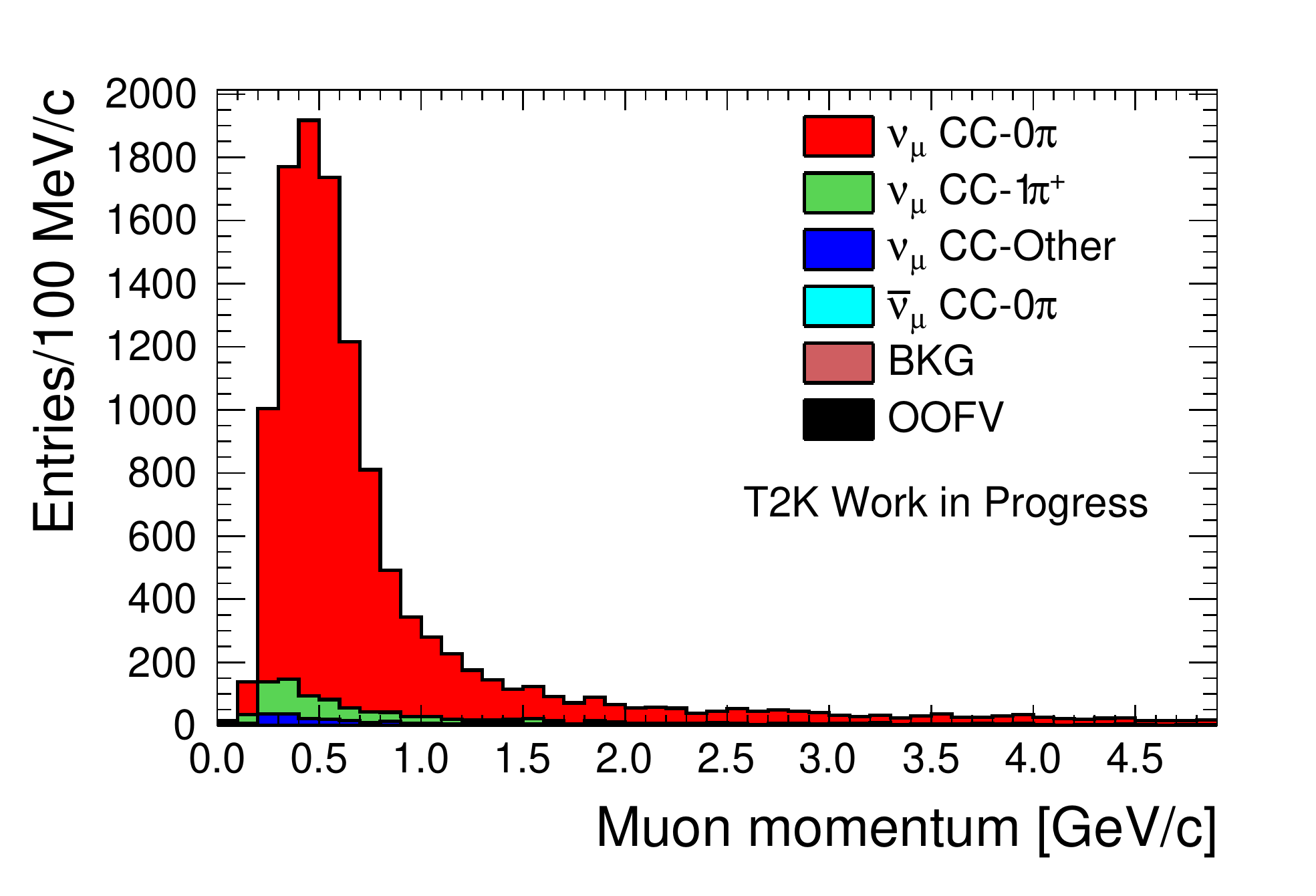}
	\includegraphics[width=.32\textwidth]{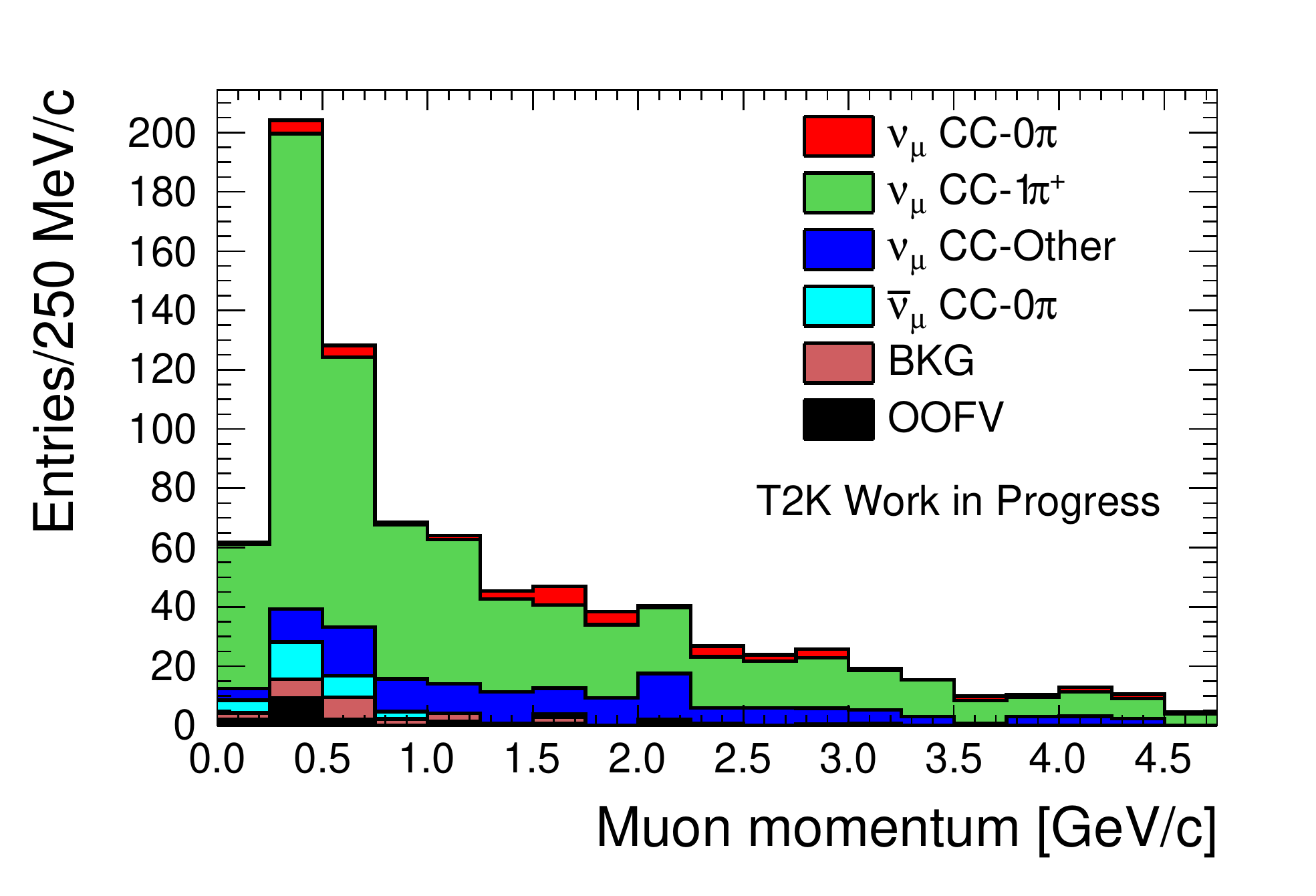}
	\includegraphics[width=.32\textwidth]{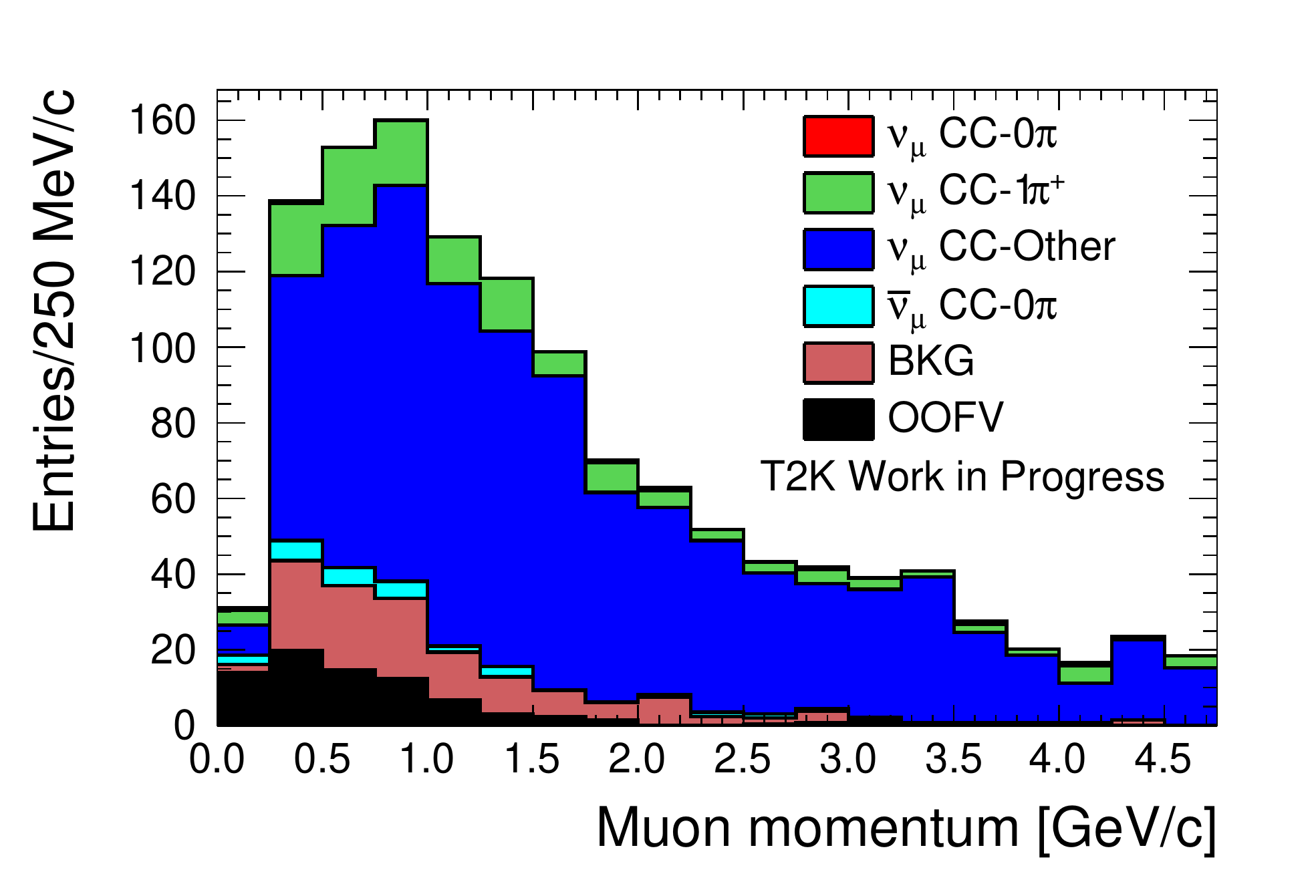}
	\caption{Muon momentum distribution for signal (top left) and control regions in the $\nu_\mu$ selection: CC-1$\pi^+$ (top right) and CC-Other (bottom center). MC is normalized to the number of proton on target (POT) in real data ($\sim$5.73$\times$10$^{20}$).}\label{numu}
\end{figure}

\begin{figure}[htbp]
	\centering
	\includegraphics[width=.35\textwidth]{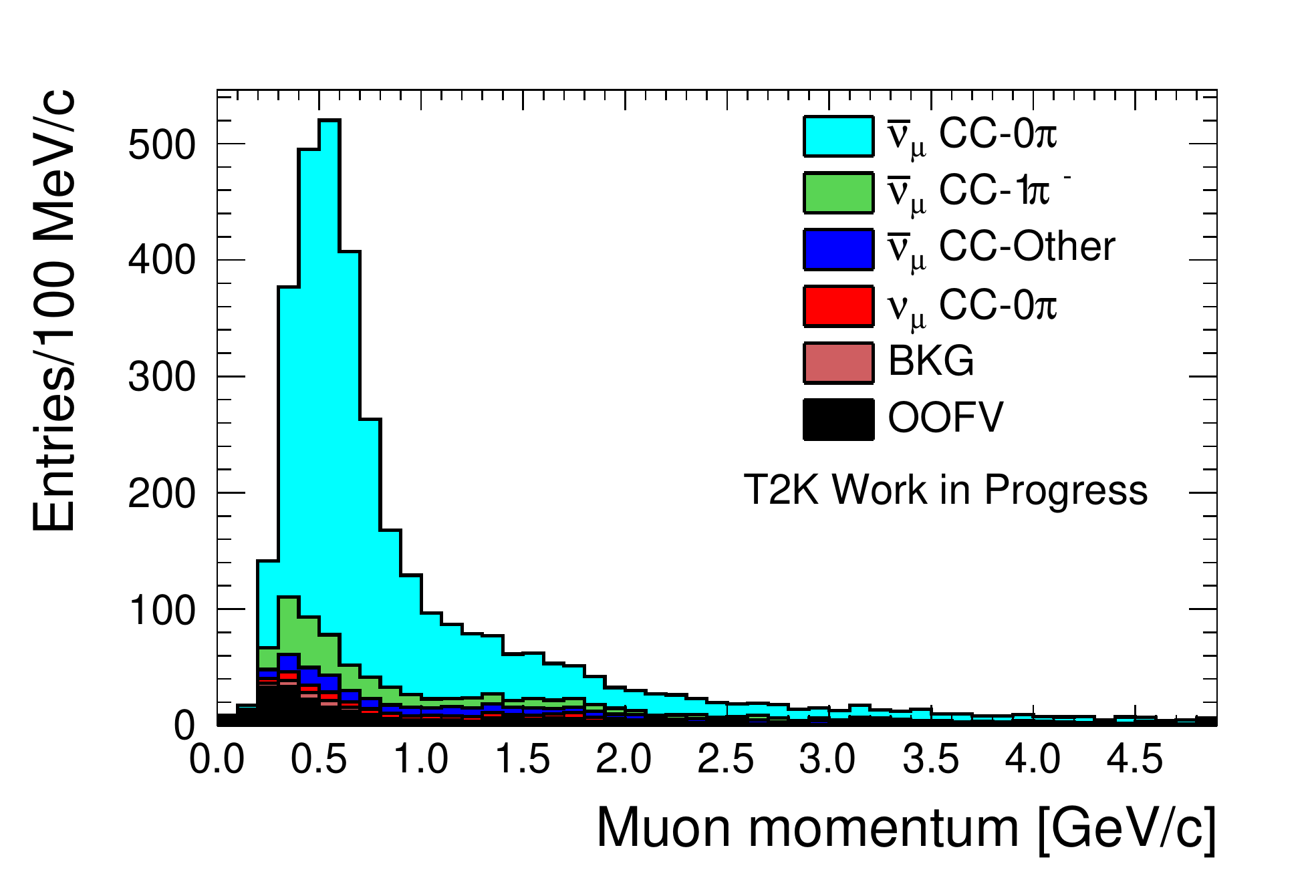}
	\includegraphics[width=.35\textwidth]{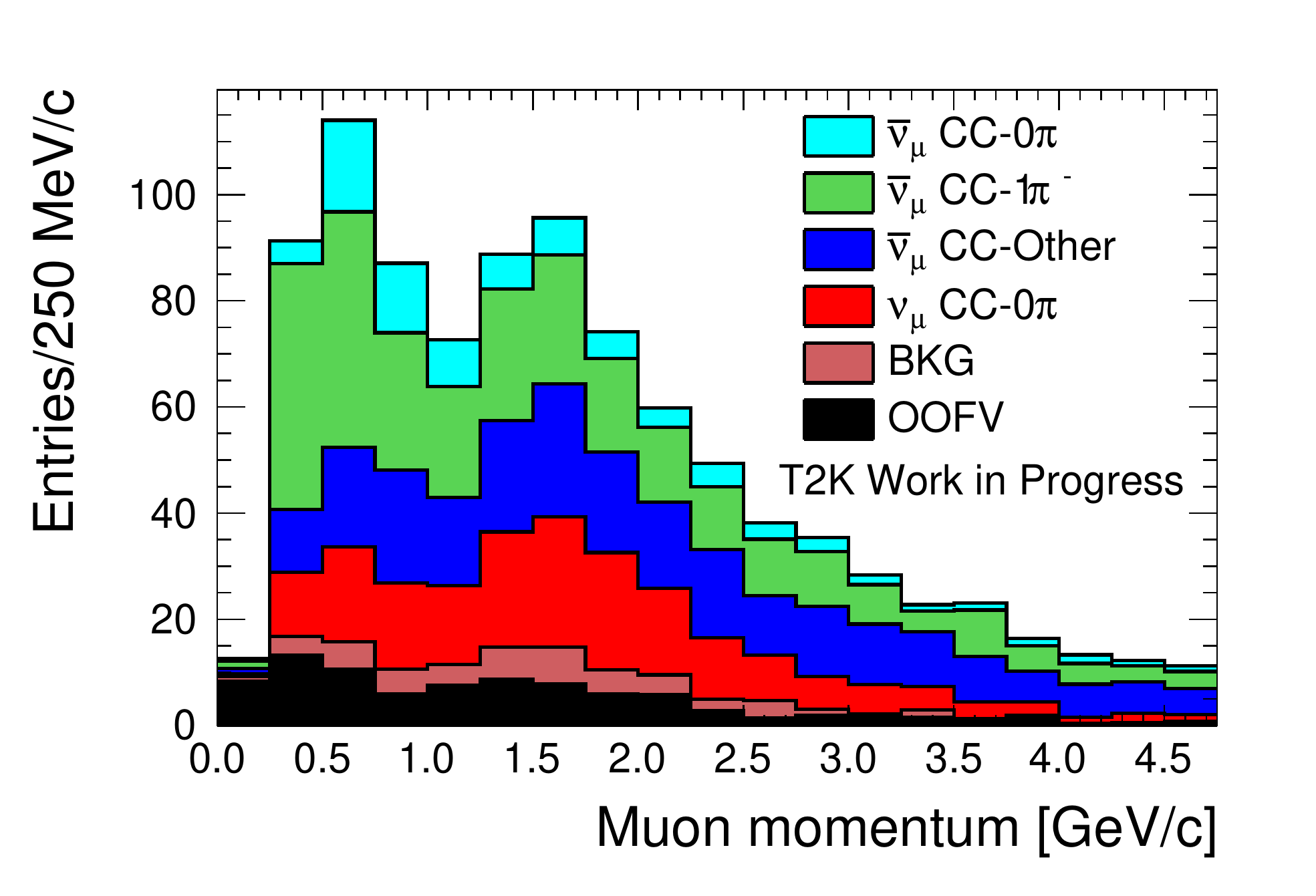}
	\caption{Muon momentum distribution for CC-1Track (left) and CC-NTracks (right) sample in the $\bar{\nu}_\mu$ selection. MC is normalized to the number of POT in real data ($\sim$3.67$\times$10$^{20}$).}\label{antinumu}
\end{figure}
\noindent
The FGD1 is used as the target and tracks are reconstructed in TPC1, if they are backward-going or in TPC2 if they are forwad-going or in the Ecal around the FGD if they have an high angle respect to the beam direction. The timing information between each subdetector is used as well to recostruct the sense of the track. The backward-going and high angle tracks are selected only in the $\nu_\mu$ selection; a similar selection for the $\bar{\nu}_\mu$ events is foreseen as future improvement. The muon candidate is identified as the highest-momentum negatively (positively) charged track which passes the TPC particle identification cut for the $\nu_\mu$ ($\bar{\nu}_\mu$) selection. The selected charged-current (CC) sample for the $\nu_\mu$ selection is split into three subsamples: CC-0$\pi$ characterized by one $\mu^-$ and any number of protons in the final state, CC-1$\pi^+$ by one $\mu^-$ and one $\pi^+$ and CC-Other by all the other events. In the $\bar{\nu}_\mu$ the CC sample is split in two subsample: CC-1Track characterized by only one $\mu^+$ and CC-NTracks with more than one track. The CC-1$\pi^+$, CC-Other and CC-NTracks subsamples act as control regions in order to extract from data the normalization and the shape of the background. The results of the selection are shown in Fig.~\ref{numu} and~\ref{antinumu} for the  $\nu_\mu$ and $\bar{\nu}_\mu$ respectively. Another future improvement to the $\bar{\nu}_\mu$ selection will be to split the CC sample in the same three subsamples selected in the $\nu_\mu$ selection in order to have the same signal and control regions definition in both the selections.

\begin{figure}[htbp]
\centering
\includegraphics[width=.35\textwidth]{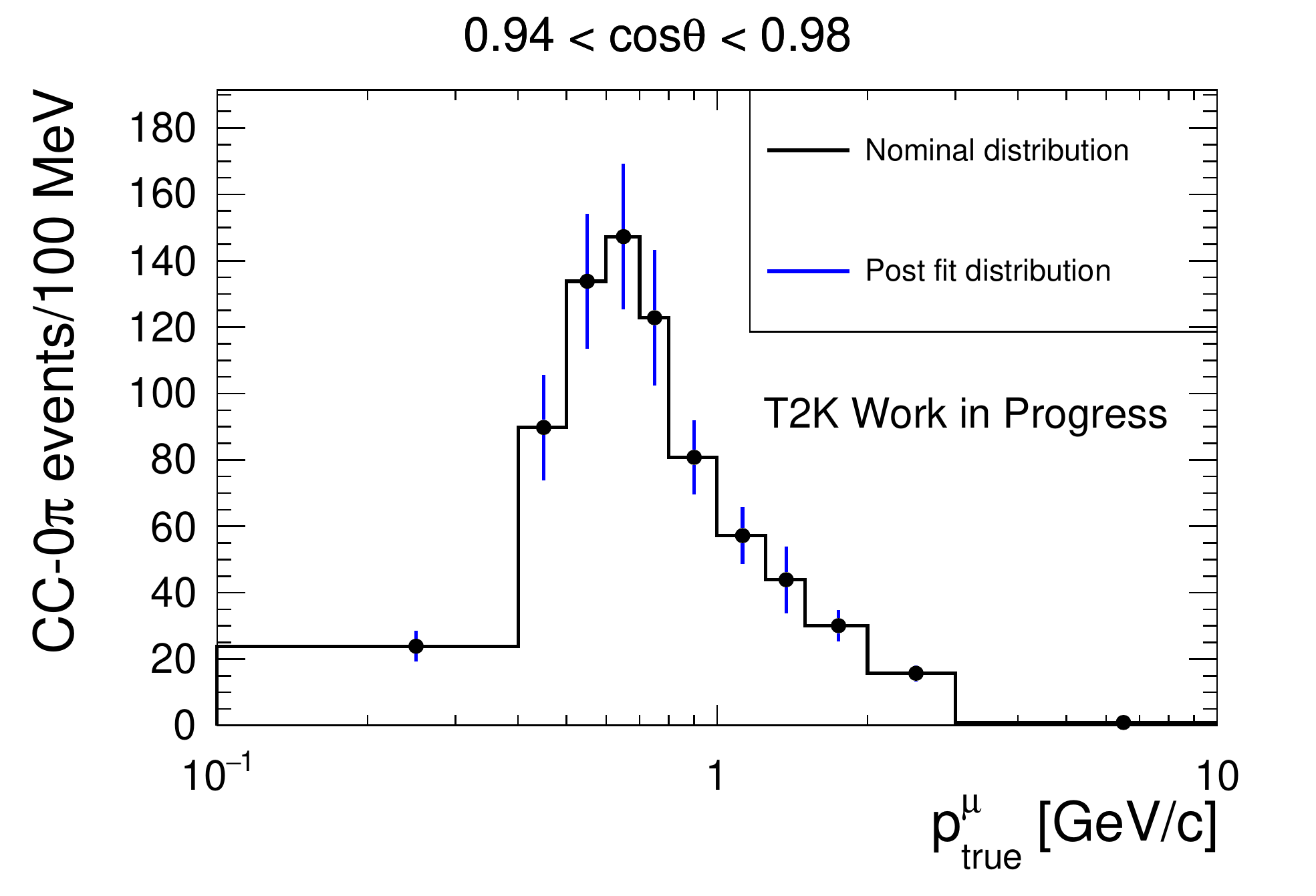}
\includegraphics[width=.35\textwidth]{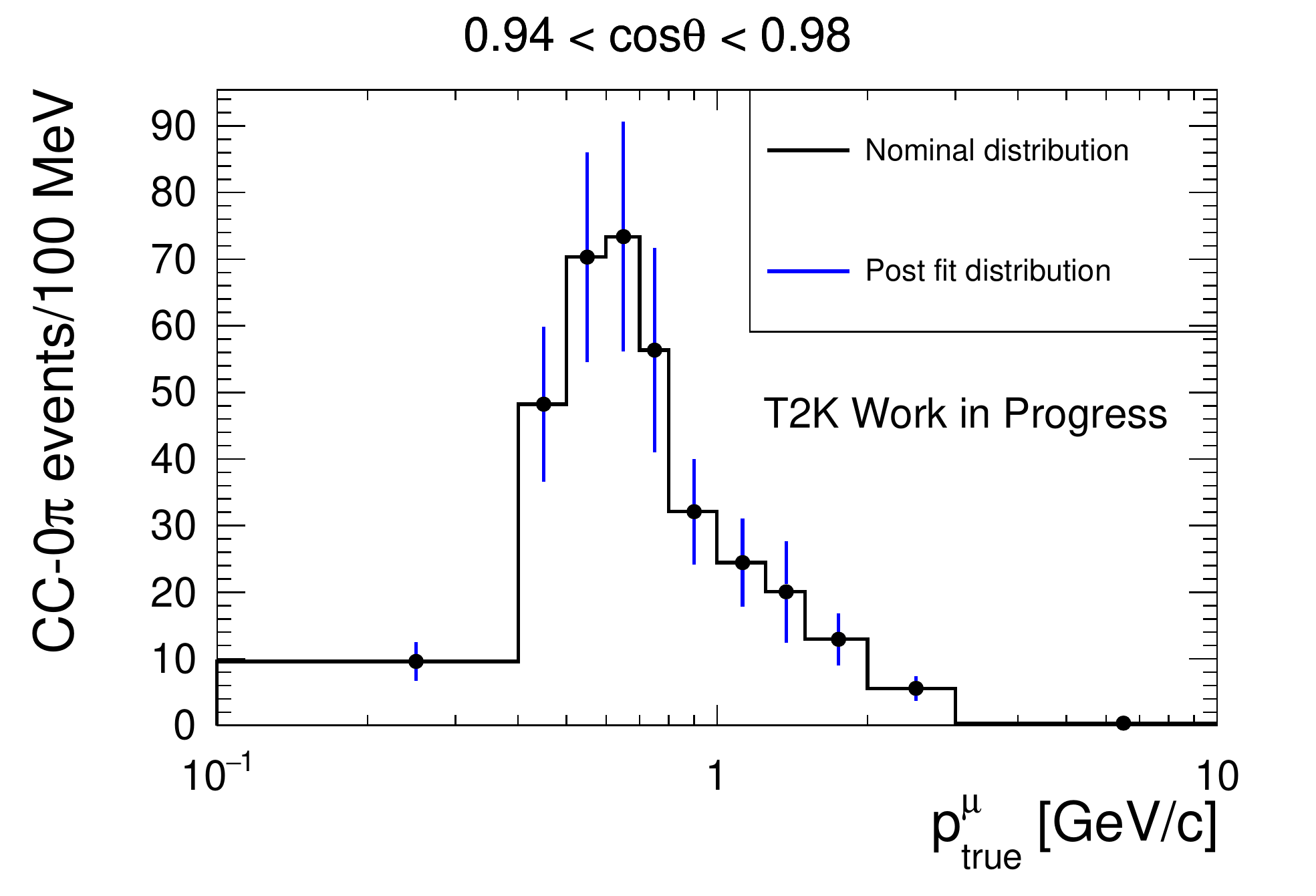}
\caption{Fit of the NEUT MC prediction with itself for one bin in cosine of the scattering angle of the outgoing muon for $\nu_\mu$ (left) and $\bar{\nu}_\mu$ (right) as a function of muon momentum with statistical error bar only.}\label{fig:asimovfit}
\end{figure}

\noindent
A preliminary result obtained fitting the NEUT MC prediciton with itself (Asimov data set) is shown in Fig.~\ref{fig:asimovfit}. The comparison between the nominal distribution and the result of the fit procedure show perfect agreemet, as expected with a well designed analysis. 

\begin{figure}[htbp]
	\centering
	\includegraphics[width=.35\textwidth]{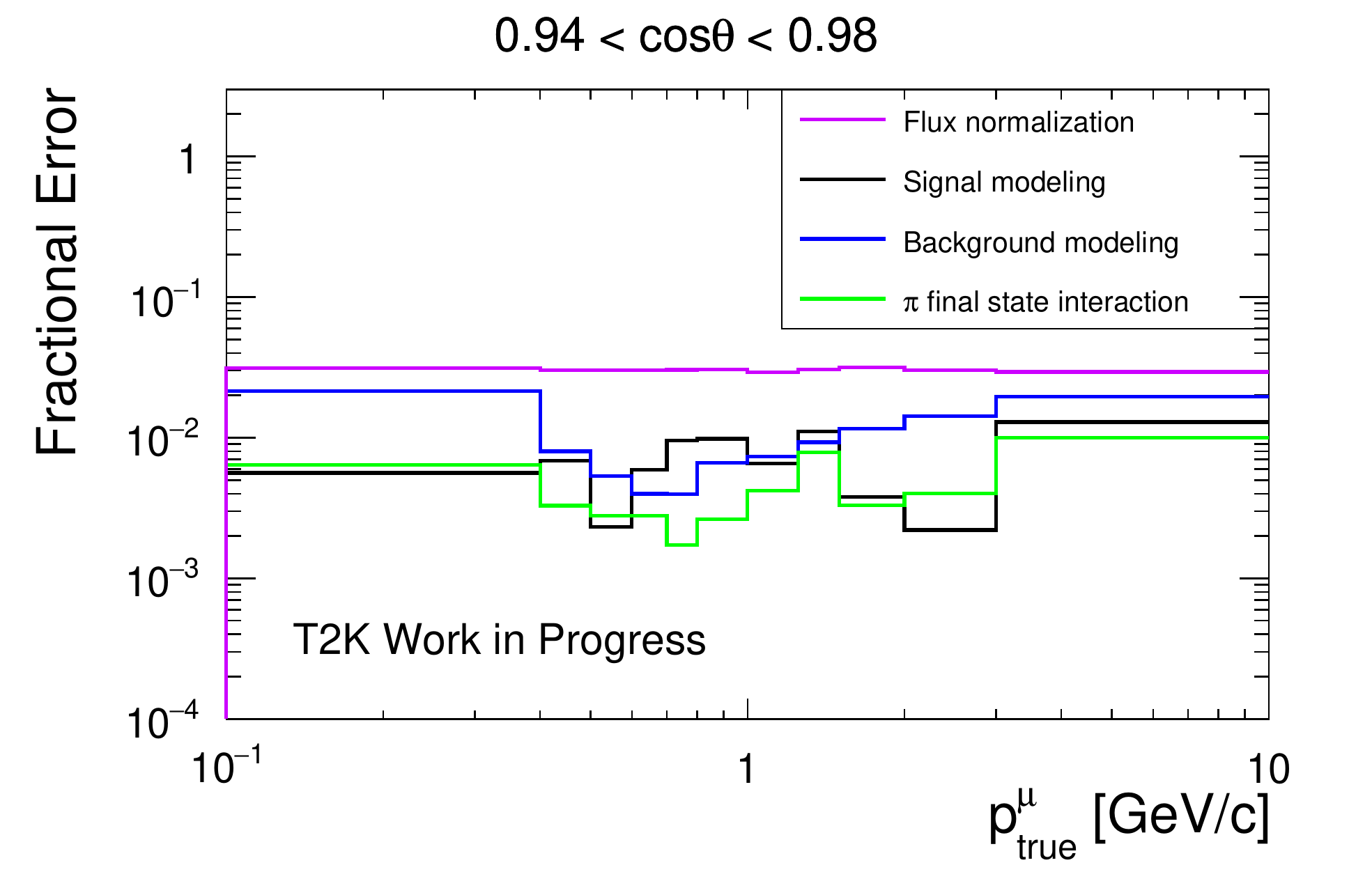}
	\includegraphics[width=.35\textwidth]{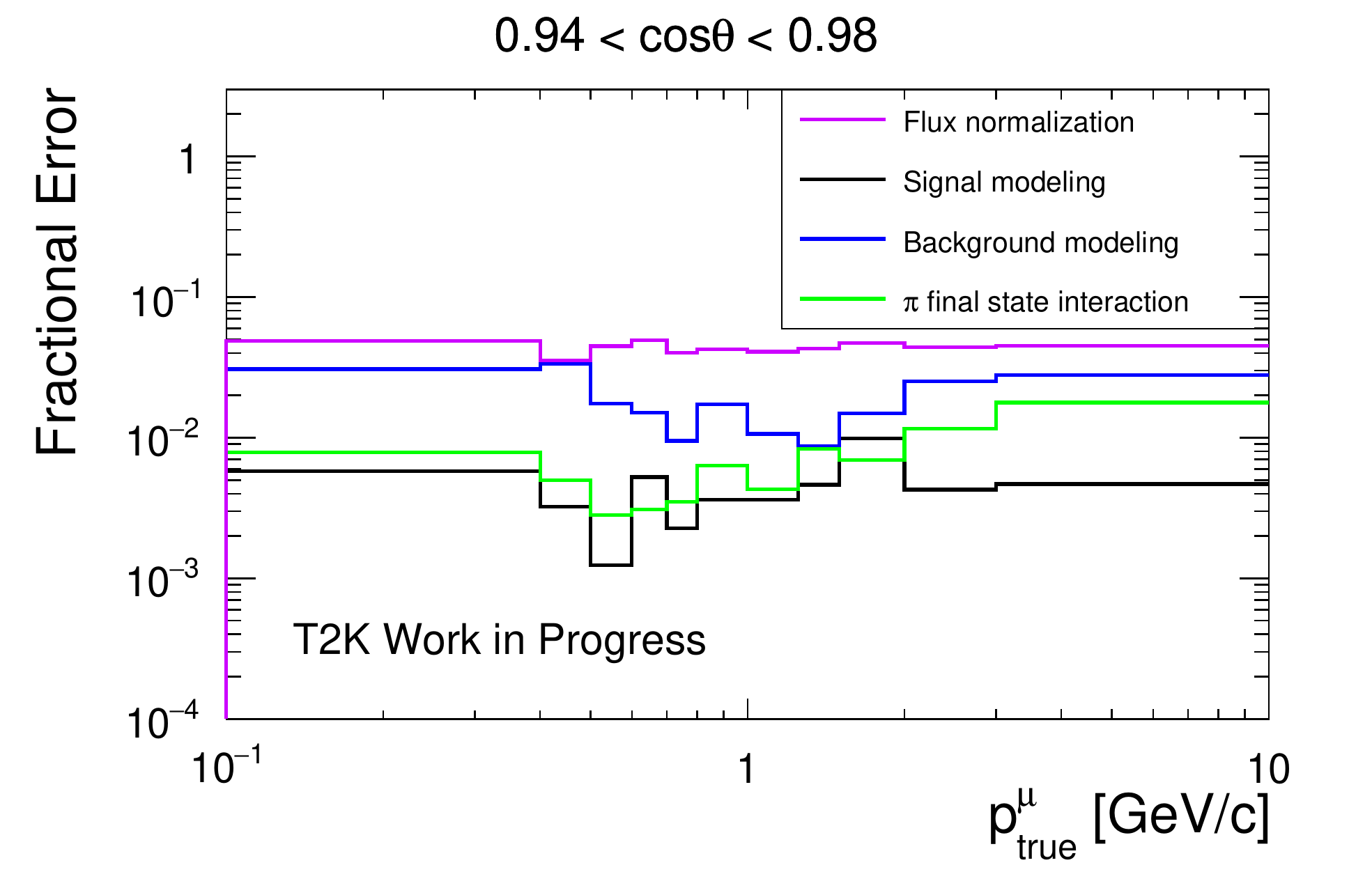}
	\caption{Preliminary evaluation of the systematic errors affecting the $\nu_\mu$ (left) and $\bar{\nu}_\mu$ (right) cross section for one bin in cosine of the scattering angle of the outgoing muon and as a funtion of muon momentum.}\label{fig:allsyst}
\end{figure}

\noindent
The flux, model and detector uncertainties are treated as nuisance parameters adding a penalty term to the likelihood and are evaluated with `toy' Monte Carlo experiments. The background model parameters are costrained by fitting the signal regions simultaneously with the control regions. A first evaluation of the systematics are shown in Fig.~\ref{fig:allsyst}, where the different colors indicates the different type of systematic. To evaluate the statistical uncertainty the number of reconstructed events in each bin is fluctuated according to the Poisson distribution.
 
\section{Conclusion}

The study of muon neutrino and antineutrino cross sections will lead to a deeper understanding of the cross section model for the CCQE scattering, vital for any oscillation analysis in long baseline experiments in the near and far future. In this work the selection of the signal regions and the control regions has been shown with a first estimation of the systematic uncertainties. The improved selection for the $\bar{\nu}_\mu$ sample is foreseen. After this improvements the systematics related with the detector will be evaluated. Along with the neutrino and antineutrino cross sections also different linear combination of the two will be evaluated.

\end{document}